 \documentclass[journal]{IEEEtran}
 \usepackage{amssymb,stmaryrd,amsmath,amsfonts,rotating,mathrsfs,TIKZ}

 \usepackage[noadjust]{cite}
 \usepackage{color}
 \usepackage{epic}

\newcommand{\aset}{{\mathcal A}}
\newcommand{\bset}{{\mathcal B}}

\newtheorem{theorem}{Theorem}

\newcommand{\btheo}{\begin{theorem}}
\newcommand{\etheo}{\end{theorem}}
\newcommand{\bproof}{\begin{proof}}
\newcommand{\eproof}{\end{proof}}
\newtheorem{definition}[theorem]{Definition}
\newcommand{\bdefi}{\begin{definition}}
\newcommand{\edefi}{\end{definition}}
\newtheorem{fact}[theorem]{Fact}
\newcommand{\bprop}{\begin{fact}}
\newcommand{\eprop}{\end{fact}}
\newtheorem{corollary}[theorem]{Corollary}
\newcommand{\bcor}{\begin{corollary}}
\newcommand{\ecor}{\end{corollary}}
\newtheorem{example}[theorem]{Example}
\newcommand{\bex}{\begin{example}}
\newcommand{\eex}{\end{example}}
\newtheorem{lemma}[theorem]{Lemma}
\newcommand{\blemma}{\begin{lemma}}
\newcommand{\elemma}{\end{lemma}}
\newtheorem{remark}[theorem]{Remark}
\newcommand{\bremark}{\begin{remark}}
\newcommand{\eremark}{\end{remark}}
\newtheorem{conj}[theorem]{Conjecture}
\newcommand{\bconj}{\begin{conj}}
\newcommand{\econj}{\end{conj}}

\newcommand{\naturals}{\ensuremath{\mathbb{N}}}

\def\0{{\tt 0}} 
\def\1{{\tt 1}} 
\def\?{{\tt *}} 
\newcommand{\dens}[1]{\mathsf{#1}}
\newcommand{\Ldens}[1]{\dens{#1}}

\newcommand{\EEx}{\hfill $\Diamond$}
\newcommand{\EDe}{\hfill $\Diamond$}

\allowdisplaybreaks
\begin{document}
\title{Universal Polar Codes}
\author{S.~Hamed~Hassani and~R\"{u}diger~Urbanke%
\thanks{S. H. Hassani and R. Urbanke are with the School of Computer and Communication Science,
EPFL, CH-1015 Lausanne, Switzerland
(e-mail: \{seyehamed.hassani, ruediger.urbanke\}@epfl.ch).}%
}
 
\maketitle
\begin{abstract}
Polar codes, invented by Arikan in 2009, are known to 
achieve the capacity of any binary-input memoryless output-symmetric
channel. Further, both the encoding and the decoding can be
accomplished in $O(N \log(N))$ real operations, where $N$ is the
blocklength.

One of the few drawbacks of the original polar code construction
is that it is not {\em universal}. This means that the code has to
be tailored to the channel if we want to transmit close to capacity.

We present two ``polar-like'' schemes which are capable of achieving
the compound capacity of the whole class of binary-input memoryless
output-symmetric channels with low complexity.

Roughly speaking, for the first scheme we stack up $N$ polar blocks
of length $N$ on top of each other but shift them with respect to
each other so that they form a ``staircase.'' Then by coding across
the columns of this staircase with a standard Reed-Solomon code,
we can achieve the compound capacity using a standard successive
decoder to process the rows (the polar codes) and in addition a
standard Reed-Solomon erasure decoder to process the columns.
Compared to standard polar codes this scheme has essentially the
same complexity per bit but a block length which is larger by a
factor $O(N \log_2(N)/\epsilon)$. Here $N$ is the required blocklength
for a standard polar code to achieve an acceptable block error
probability for a single channel at a distance of at most $\epsilon$
from capacity.

For the second scheme we first show how to construct a true polar
code which achieves the compound capacity for a finite number of
channels.  We achieve this by introducing special ``polarization''
steps which ``align'' the good indices for the various channels.
We then show how to exploit the compactness of the space of
binary-input memoryless output-symmetric channels to reduce the
compound capacity problem for this class to a compound capacity
problem for a finite set of channels. This scheme is similar in
spirit to standard polar codes, but the price for universality is
a considerably larger blocklength.

We close with what we consider to be some interesting open problems.
\end{abstract}

\section{Introduction} Consider a communication scenario where the
transmitter does not know the channel over which transmission takes
place but only has knowledge of a {\em set} that the actual channel
belongs to.  Hence we require that the coding scheme must be reliable
for every channel in this set.  The preceding setup is known as the
{\em compound channel} scenario and the maximum achievable rate is
known as the {\em compound capacity}.  Several variations on this
theme are possible and useful. We consider the case where the
transmitter only has knowledge of the set but the receiver knows
the actual channel that was used. This is not unrealistic. If the
channel is constant or changes very slowly then the receiver has
ample of time and data to estimate the channel very accurately.

Let $\mathcal{W}$ denote the set of channels.  The compound capacity
of $\mathcal{W}$, denote it by $C(\mathcal{W})$, is defined as the
maximum rate at which we can reliably transmit irrespective of which
channel from $\mathcal{W}$ is chosen.  It was shown in \cite{BBT59}
that
\begin{align} \label{compound_best} 
C(\mathcal{W}) = \max_{Q}\inf_{\Ldens{a}\in
\mathcal{W}}I_Q(\Ldens{a}), 
\end{align} 
where $I_Q(\Ldens{a})$ denotes the mutual information between the
input and the output of $\Ldens{a}$, with the input distribution
being $Q$.  Note that the compound capacity of $\mathcal{W}$ can
be strictly smaller than the infimum of the individual capacities.
This happens only if the capacity-achieving input distributions
for the individual channels are different.

We restrict our attention to the class of binary-input memoryless
output-symmetric (BMS) channels.  As the capacity-achieving input
distribution for all BMS channels is the uniform one (and hence in
particular the same), it follows that for any collection $\mathcal{W}$
of BMS channels  the compound capacity is equal to the infimum of
the individual capacities.

Why is this problem of practical relevance?  When we design a
communications system we typically start with a mathematical model.
But in reality no channel is exactly equal to the assumed model.
Depending on the conditions of the transmission medium, the channel
will show some variations and deviations. Therefore, designing
low-complexity universal coding schemes  is a natural and important
problem for real systems.  Spatially coupled codes \cite{KRU12}
were the first class of low complexity codes to be shown to be
universal.

Consider standard polar codes  with the standard successive decoder
\cite{Ari09}.  For this scheme the question of universality was
addressed in \cite{HKU09}.  By deriving a sequence of upper and
lower bounds, it was shown that in general the compound capacity
under successive decoding is strictly smaller than the unrestricted
compound capacity described in \eqref{compound_best}.  In words,
standard polar codes under successive decoding are not universal.

One might wonder if this lack of universality is due to the code
structure or due to the (suboptimal) successive decoding procedure.
To answer this question, let us consider polar codes under MAP
decoding. Let $C \in [0,1]$ and consider the polar code (with the
standard kernel $G_2 = \bigl(\begin{smallmatrix} 1 & 0\\ 1&1
\end{smallmatrix} \bigr)$) designed for the binary symmetric 
channel (BSC) with capacity $C$. It is shown in \cite{erenthesis}
that under MAP decoding such a code achieves the compound 
capacity if we take ${\mathcal W}$ to be the class of BMS  channels
of capacity $C$.  Consequently, polar codes, decoded with the
optimal MAP decoder, are universal.  Hence, it is the suboptimal 
decoder that is to fault for the lack of universality.  

It is therefore interesting to ask whether some suitable modification
of the standard polar coding scheme allows us to construct
``polar-like'' codes which are universal under low-complexity
decoding. As we will show, the answer is yes. In fact, we present
two solutions. The first solution combines polar codes with
Reed-Solomon (RS) codes which are optimal for the (symbol) erasure
channel\footnote{Note that \cite{BJE10, MELK113} proposes a concatenated
coding scheme involving polar codes and RS codes.  The idea for
such a scheme is to mitigate the effect of error propagation of the
successive decoder by making use of the error protection capabilities
of the RS code.}.  The second solution is a slight modification of
the standard polar coding scheme and it is itself a polar code where
channels are combined in a specific way in order to guarantee
universality.

In independent work \c{S}a\c{s}o\u{g}lu and Wang also consider the
problem of constructing universal polar codes.  Their solution, see
\cite{SaW13}, is based on introducing two types of polarization
steps.  The first one is the usual polarization step and it is used
to achieve a low error probability. The second one, which is novel,
guarantees that the resulting code is universal.

Before we present our schemes let us agree on notation
and let us recall some facts. 

Consider a standard polar block of length $N=2^n$ generated by the
matrix $G_2$.  Note that we use the word {\em block} to denote the
structure implied by the $n$-fold Kronecker product of $G_2$,
together with the implied decoding order of the successive decoder.

As a next component we need to specify the channel over which
transmission takes place.  All channels we consider are BMS channels.
Assume that we are given the channel $\Ldens{a}$, where $\Ldens{a}$
might be a binary-erasure channel (BEC), a binary symmetric channel
(BSC), a binary additive white-Gaussian noise channel (BAWGNC) or
any other element of the class of BMS channels.  We denote its
capacity by $C(\Ldens{a})$. Once we are given the channel we can
compute for the given length $N$ the set of ``good'' polar indices.
Call this set $\aset$.  There are many possible ways of defining
this set. To be concrete, we will use the following convention.
Fix the rate $R$ where $0 < R < C(\Ldens{a})$.
Compute for the given channel the Battacharyya constants associated
to all indices $i$, $1 \leq i \leq N$. Sort these numbers from
smallest to largest.  Include in $\aset$ the smallest $R N$ such
indices.  Note that the sum of the Battacharyya constants of the
included indices is an upper bound on the block error probability
under successive decoding. We denote this error probability by $P(\Ldens{a})$.  Efficient
algorithms to determine the set of good indices can be found in
\cite{TV, RHTT}.

In this respect the following fact, first stated in \cite{ArT09},
is important: For any BMS channel $\Ldens{a}$, any $0< R <
C(\Ldens{a})$, and any $0 < \beta < \frac12$, we have $P(\Ldens{a})
\leq  c(C(\Ldens{a}), R, \beta) 2^{-N^\beta}$, where $c(C(\Ldens{a}),
R, \beta)$ only depends on $C(\Ldens{a})$, the chosen rate $R$, and
$\beta$, but is universal with respect to $\Ldens{a}$.   This means
in particular that for any fixed $k>0$, by choosing $N$ sufficiently
large, we can make $N^k P(\Ldens{a})$ as small as desired.  Note
that this bound not only holds for the block error probability
but also for the sum of the Battacharyya constants of the included
channels.

In the previous paragraphs we used the word ``index'' to refer to
one of the synthetic channels which are created by the polarization
process. The reason for using ``index'' and not ``channel'' or
``synthetic channel'' is that we consider transmission
over a set of channels ${\mathcal W}$ and hence referring to both, the
actual transmission channel and the synthetic channels created by
the polarization process, as channels might lead to confusion.  In
the sequel we will always assume that these indices are labeled
from $1$ to $N$ and that the processing order of the successive
decoder is the one implied by this labeling (i.e., we first process
index $1$, then $2$, and so on).

We will also need a universal upper bound on the
blocklength which is required if we want to transmit with a standard
polar code close to capacity. Such a bound is stated in the next lemma.
\begin{lemma}[Universal Upper Bound on Block Length -- \cite{HAU13}]\label{lem:uniform}
For $0<C<1$, $\Delta>0$, and $P>0$ define
\begin{align*}
n(C, \Delta, P) = \lceil 7 \log_2 \frac{1}{\Delta} + c(C, P) (\log_2(\log_2 \frac{4}{\Delta}))^2 \rceil.
\end{align*}
Then a polar code of length $N \geq 2^{n(C, \Delta, P)}$ and rate
$R=C-\Delta$ designed for $\Ldens{a} \in \text{BMS}(C)$ has a block
error probability under successive decoding of at most $P/N^2$.
Here, $c(C, P)$ only depends on $C$ and $P$ but is
independent of $R$, and $\Ldens{a}$.  
\end{lemma}

{\em Discussion:} The scaling of $P/N^2$ is somewhat arbitrary. The
same result, albeit with a different constant, is true for the more
general case where we require $P/N^k$, $k >0$.

As a final notational convention, we will write $\text{BMS}(C)$ to
denote the set of all BMS channels of capacity at least $C$.

\section{Base Scheme for Two Channels}\label{sec:basescheme} 
Consider two channels, call them $\Ldens{a}$ and $\Ldens{b}$, both
of capacity $C$.  This means that ${\mathcal W}=\{\Ldens{a}, \Ldens{b}
\}$.  Assuming that both channels have capacity $C$ entails no
essential loss of generality since for the class of BMS channels
the compound capacity of a set of channels is equal to the minimum
of the capacities, as was mentioned in the introduction.

Consider two polar blocks of length $N$ and let $\aset_N$ and
$\bset_N$ be the set of ``good'' indices for channel $\Ldens{a}$
and $\Ldens{b}$, respectively. What we mean with this is that with
this chosen set we get ``acceptable'' block error probabilities,
call them $P(\Ldens{a})$ and $P(\Ldens{b})$, respectively. As we have
discussed in the introduction, one convenient way of defining this set
is to fix a rate $0<R<C$ and then to include the $NR$ indices of the block
of length $N$ that have the smallest Battacharyya parameters.

Since by Lemma~\ref{lem:uniform} polar codes achieve the capacity
uniformly over the class of BMS channels, it entails further no
essential loss of generality if we assume that $|\aset_N|=|\bset_N|$.

The most obvious way of constructing a polar code for this compound
case is to place the information in the set $\aset_N \cap \bset_N$,
i.e., to place information only in the indices which are good for
{\em both} channels. The block error probability under the standard
successive decoder is in this case bounded above by $\max\{P(\Ldens{a}),
P(\Ldens{b})\}$, which is good news.  Unfortunately, as was mentioned
in the introduction, it was shown in \cite{HKU09} that such a scheme
in general results in rates which are strictly below the
compound capacity even if we let $N$ tend to infinity.  This means
that for large $N$, $|\aset_N \cap \bset_N|/N \leq \alpha
\min\{|\aset_N|, |\bset_N|\} / N$, where $\alpha<1$.  One notable
exception is the case where the channels are ordered by degradation,
but this covers only a small range of cases of interest, see
\cite{Kor09thesis}.

\section{Capacity Gap}
Consider again the case ${\mathcal W}=\{\Ldens{a}, \Ldens{b}\}$ and
let $0< P < 1$.  Assume that we include in the set $\aset_N$ the
maximum number of indices so that the sum of their Battacharyya
constant (with respect to channel $\Ldens{a}$) does not exceed $P$
and that we define $\bset_N$ in the equivalent manner.  Then
we know from \cite{Ari09} that the limits $\lim_{N \rightarrow
\infty} \frac{|\aset_N|}{N}$ and $\lim_{N \rightarrow \infty}
\frac{|\bset_N|}{N}$ exist and are equal to $C(\Ldens{a})$ and
$C(\Ldens{b})$ respectively. Further, it was shown in \cite{HKU09}
that \begin{align*} 
\lim_{N \rightarrow \infty} \frac{|\aset_N \cap \bset_N|}{N}
\end{align*} exists. Call this limit $C(\Ldens{a}
\cap \Ldens{b})$.  More generally, we can define the limit
$C(\cap_{\Ldens{a} \in {\mathcal W}} \Ldens{a})$.  This is the rate
which we can achieve if we only transmit on those indices which are
good for all channels in ${\mathcal W}$.  We can now define the
{\em gap}, call it $\Delta(\cap_{\Ldens{a} \in {\mathcal W}} \Ldens{a})$, as  $\Delta
(\cap_{\Ldens{a} \in {\mathcal W}} \Ldens{a}) = \min_{\Ldens{a} \in
{\mathcal W}} C(\Ldens{a})- C(\cap_{\Ldens{a} \in {\mathcal W}}
\Ldens{a})$. For convenience of notation let us define $C({\mathcal
W})=\min_{\Ldens{a} \in {\mathcal W}} C(\Ldens{a})$ as a shorthand
for the compound capacity.

\section{Scheme I}\label{sec:finitesettwo}
\begin{figure}[ht!]
\begin{center}
\input{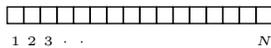}
\end{center}
\caption{A polar block of length $N$ as a row vector.}\label{fig:basicpolarblock}
\end{figure}

\begin{figure}[ht!]
\begin{center}
\input{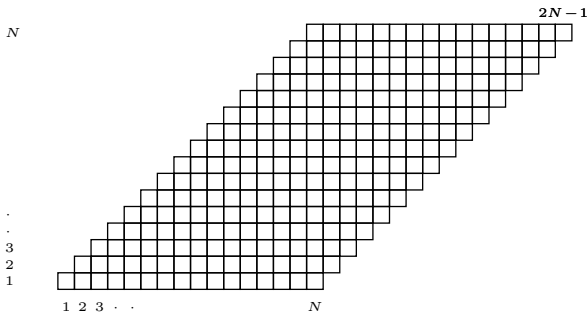}
\end{center}
\caption{A {\em staircase} consisting
of $N$ basic polar blocks stacked on top of each other, where the
$j$-th such block (counted from the bottom), $1 \leq j \leq N$, is
shifted $(j-1)$ positions to the right. The columns are labeled from
$1$ to $2N-1$ and the rows are labeled from $1$ to
$N$.}\label{fig:staircase}
\end{figure}

Let us now describe our first scheme. Represent a polar block of
length $N$ by a row vector as in Figure~\ref{fig:basicpolarblock}.
Take $N$ such blocks and construct a {\em staircase} by
stacking these blocks on top of each other as shown in
Figure~\ref{fig:staircase}. Note that the $j$-th such block (counted
from the bottom), $1 \leq j \leq N$, is shifted $(j-1)$ positions
to the right. 

\begin{figure}[ht!] 
\begin{center} 
\input{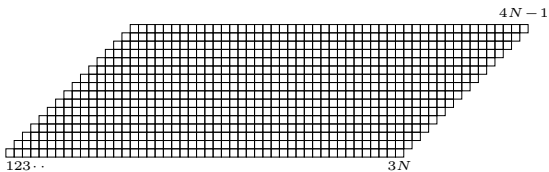}
\end{center} \caption{An {\em extended staircase}
for $N=16$ and $k=3$.}\label{fig:extendedstaircase} 
\end{figure}

Next, extend the staircase by placing $k$ copies of this staircase
horizontally next to each other in a consecutive manner, where $k
\in \naturals$ is a parameter of the construction. Call the result
an {\em extended} staircase. This is shown in Figure~\ref{fig:extendedstaircase}
for $N=16$ and $k=3$.

\begin{figure}[ht!] 
\begin{center} 
\input{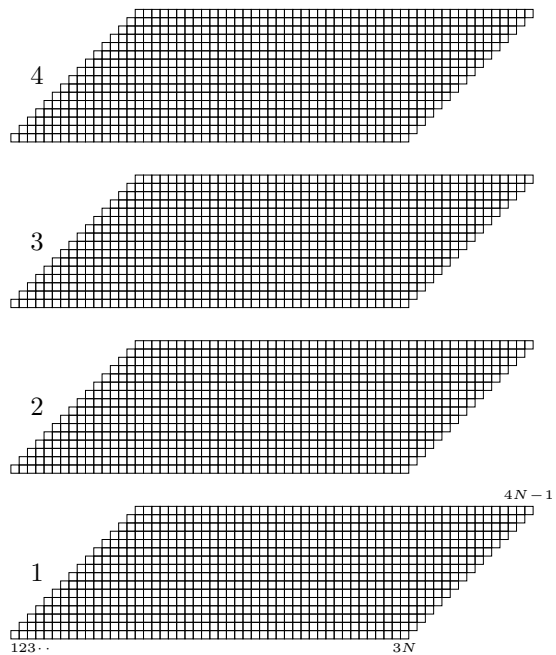}
\end{center} \caption{The scheme for
$N=16$ and $k=3$, consisting of $\log_2(16)=4$ extended staircases.}\label{fig:schemeone} 
\end{figure}

Finally, take $\log_2(N)=n$ such extended staircases.  Graphically
we think of them as being placed in a vertical direction on top of
each other.  Figure~\ref{fig:schemeone} shows the result for
$N=16$ and $k=3$.  

It remains to explain where to place information and how to recover
it.  Note that each extended staircase has width $(k+1) N-1$, and
we assume that the (column) indices run from $1$ to $(k+1) N-1$.
Note further that, except for the boundaries, the columns of each
extended staircase have height $N$. More precisely, all columns in
the range $N \leq i \leq k N$ have height $N$. We say that such a
column has {\em full} height.  As a final observation, note that
in a column of full height, due to the shifts, we ``see'' exactly
the same indices (channels) as in a standard polar block of length
$N$. In other words, we can think of one column of full height as
(a cyclic shift of)
 a standard polar block. This is one key reason why our construction
 works.

Now recall that according to Lemma~\ref{lem:uniform}, regardless of
what channel from $\text{BMS}(C)$ is chosen, for sufficiently large
$N$, the number of good indices in one polar block is very close
to $N C$ and the notion of ``very close'' is uniform with respect
to the channel. In words, regardless of what channel is chosen,
out of the $N$ indices in a column about $NC$ can be (correctly) decoded  and
they are decoded with high probability. Further, since we know at
the receiver what channel has been used, we know which of the indices can
be decoded.  Therefore, we can treat the undecoded indices as {\em
erasures.}

The idea is therefore simple. Use in each full-height column an
erasure code so that we can reconstruct the whole column if we know
roughly $N C$ components of it. Since we want to do this without loss,
we wish to use a {\em maximum distance separable} (MDS) code. Since
binary MDS codes only exist for very few parameters we take 
$\log_2(N)=n$ such staircases. Exploiting this fact we can code
over $\text{GF}(N)$, and over this field there do exist MDS codes
of any dimension up to length $N$, namely RS codes. Hence, the idea is 
to use a RS code for each column and then the resulting vector in each row is further encoded 
using the polar transform. 

Let us explain this in more detail. In order not to complicate
things we first assume that we have at our disposal binary MDS codes
of length $N$ and dimension a little bit smaller than $N C$.  In
this case a single extended staircase suffices. Let us explain how
we encode in this case.

Recall how encoding is done for a standard polar code. In this case
we first designate which of the $N$ positions carry information and
which ones are frozen. We then load the information positions and
place a known pattern in the remaining (frozen) positions. Typically,
for convenience, this known pattern is the all zero pattern but any
pattern is possible as long as it is known at the receiver.  This
procedure gives us a vector of length $N$. To get from this vector
the codeword, we multiply the vector by the polar matrix. In other
words, we first create a vector of length $N$ which contains $N R$
information bits.  Then we transform this vector.

In the same manner the encoding process for our construction has
the same two steps.  We first fill in every element of the extended
staircase with binary symbols.  In a second step we then take each
of the polar blocks of the extended staircase and we multiply the
vector contained in this block by the polar matrix.  The final
result is our codeword. 

It remains to explain how we fill the elements of the extended
staircase.  At the boundary, i.e., in columns which are not of full
height, we fill those indices which are good indices for {\em all}
channels in the given class with information, and all other indices
are filled with a known pattern (e.g., the all-zero pattern). For
full-height columns we proceed in a different way. For each such
column take slightly fewer than $N C$ information bits and encode these bits
into a codeword of the MDS code of length $N$.  Fill in this vector
of length $N$ into this full-height column.  Repeat this procedure
for every full-height column.

Now let us look at the receiver and the decoding process.  The
decoding proceeds left to right. At time $i$ we decode all the
positions which correspond to column $i$, i.e., we can imagine that
there are $N$ polar decoders running in parallel, one on each row
of the extended staircase, but they are synchronized so that they
are all working on the same column at one point in time.

At the receiver we know the channel and hence each decoder knows
whether the index she is currently working on belongs to the good
set for this channel.  Those decoders that ``see'' a good index at
the current point in time decode this index using one more step of
the successive decoder. With high probability they will be able to
decode correctly.

How many decoders will ``see'' good indices? By construction the
proportion will be close to $N C$. This means that we can recover
reliably about $NC$ of the $N$ bits in the current column.  Now we
exploit the fact that the bits of this column form the codeword of
an MDS code of rate just a little bit below $NC$.  We can therefore
recover all the bits of this column by completing this codeword.
Performing this operation column by column we can recover the whole
extended staircase.

At the boundaries we proceed in a simpler fashion since there we
only store information in indices which are good for all channels
and all other indices are frozen.

In the above paragraph we have crucially used the following property
of polar codes. For BMS channels we can choose the value of frozen
bits in any manner we wish as long as these values are known at the
decoder.

Let us now clarify why in general we do not use only a single
extended staircase but $n$ of them. This slight modification allows
us to deal with the fact that we cannot code over the binary field
but need to code at least over a field of size $N$ in order to
construct an MDS code of length $N$ of dimension roughly $N C$.

How can we use the $\log_2(N)$ copies? The crucial observation is
that the $\log_2(N)$ copies behave essentially identical.  Therefore,
fix a particular full-height column. Assume that for a particular
channel $\Ldens{a}$ we know that lets say index $i$, $1 \leq i \leq
N$, is good. Then this index is good for all the $\log_2(N)$ extended
staircases and with high probability we recover all $\log_2(N)$ of
them. So if we think of these $\log_2(N)$ bits as one symbol of
$\text{GF}(N)$ then we can assume that this symbol is known.
Conversely, assume that this index is not good for the chosen
channel. Then it is not good for any of the $\log_2(N)$ extended
staircases and this fact is known at the receiver. So, if again we
combine these $\log_2(N)$ bits into an element of of $\text{GF}(N)$
then we can think of this element as an erasure and the overall
erasure probability is very close to $1-C$, as it should be.

Let us summarize. For columns $1 \leq i \leq N-1$ and $kN \leq i
\leq (k+1)N-1$ we load information only into those polar indices
which belong to $C(\cap_{\Ldens{a} \in \text{BMS}(C)} \Ldens{a})$.
For all other columns, i.e., the columns $N \leq i \leq k N$ we
load into the $\log_2(N)$ columns of the $\log_2(N)$ extended
staircases at position $i$ one RS codeword of length $N$ over the
field $\text{GF}(N)$, where the RS code has has rate just a little
bit less than $C$.
We then multiply each row by the polar matrix.
This specifies the encoding operation.

For the decoding, we run $N \log_2(N)$ successive decoders in
parallel, each working on one of the $N \log_2(N)$ rows of the
scheme. These decoders are synchronized in the sense that they are
processing the bits in the same column of the scheme at the same
time. Regardless of what channel the transmission takes place,
according to Lemma~\ref{lem:uniform} we can decode about a fraction
$N C$ of the $N$ positions in each extended staircase.  Therefore,
the RS code which has a rate just a little bit below $N C$ will be
able to recover all symbols.

The subsequent lemma summarizes our observations and gives the
precise parameters and the resulting bounds on the error probability
as well as the complexity.

\begin{lemma}[Universal Polar Codes]
Let $\text{BMS}(C)$ denote the set of BMS channels of capacity at
least $C$. Let $\epsilon>0$ be the allowed gap to the compound
capacity and let $P>0$ be the allowed block error probability.
Consider the above construction with the following parameters.
\begin{itemize}
\item Pick $k=\frac{2}{\epsilon}$.
\item Let $N = 2^{n(P \frac{\epsilon}{2}, \epsilon/2)}$, where $n(P, \epsilon)$ is given
in Lemma~\ref{lem:uniform}.
\end{itemize}
{\em Encoding:} Assume that for the columns $1 \leq i <N$ and $k N
< i \leq (k+1)N-1$ we load only the indices which are good for all
the channels in BMS($C$). For full-height columns, i.e., columns
with $N \leq i \leq k N$, we load the columns with RS codewords of
length $N$ over the field $\text{GF}(N)$ and of dimension $(C-\frac12
\epsilon)N$.  We then multiply each polar block by the polar matrix
to accomplish the encoding. \\
{\em Decoding:} At the decoder we proceed as follows.
For columns $1 \leq i \leq N-1$ and $kN \leq i \leq (k+1)N-1$ we
use the standard successive polar decoder to recover those indices
which are good for all channels.  For the columns $N \leq i < kN $
we first use successive polar decoding for all those rows which the
index at the intersection of this row and the current column (the
$i$-th column) is a good index for the channel at hand. This knowledge
is present at the decoder since we assume that the receiver knows
the channel over which transmission takes place and it can hence
compute these indices. We then perform a RS erasure decoder along
the column to fill in all missing information.

This results in a scheme with the following parameters
which hold uniformly over the whole class $\text{BMS}(C)$.
\begin{itemize}
\item $R \geq C(1-\epsilon)$

\item The blocklength is $N^2 \log_2(N) \frac{2}{\epsilon}$.

\item The block error probability is upper bounded by $\frac{P \log_2(N)}{N} \leq P$ uniformly over the set
$\text{BMS}(C)$.

\item The {\em encoding} complexity {\em per bit} is
$O((\log_2(N))^{\log_2(3)})$ binary operations.

\item The {\em decoding complexity} {\em per bit} is $O(\log_2(N))$
real operations (for the polar decoder) and $O((\log_2(N))^{1+\log_2(3)})$ binary
operations for the decoding of the RS code.
\end{itemize}
\end{lemma}
\begin{IEEEproof}
Let us go over each of these claims one by one.
\begin{itemize}
\item {[$R \geq C(1-\epsilon)$.]} This follows by construction.  We
loose at most a factor $(1-\epsilon/2)$ compared to the compound
capacity due to boundary effects of the staircase and a further
such factor due to the fact that we chose a finite value for $N$
and so we are bounded away from capacity.

\item {[The blocklength is $N^2 \log_2(N) \frac{2}{\epsilon}$.]}
By construction each extended staircase contains $N k$ blocks and
we have $\log_2(N)$ of those. The claim now follows by our choice
$k=\frac{2}{\epsilon}$.

\item {[The block error probability is upper bounded by $\frac{P
\log_2(N)}{N} \leq P$ uniformly over the set.]} Note that we have
in total $N \log_2(N) \frac{2}{\epsilon}$ polar blocks. By construction,
the block error probability for each of them under successive
decoding is at most $P \frac{\epsilon}{2 N^2}$ and this bound is uniform
over all BMS$(C)$ channels. The claim therefore follows by an
application of the  union bound.

\item {[The {\em encoding} complexity {\em per bit} is
$O((\log_2(N))^{\log_2(3)})$ binary operations.]} The encoding
complexity consists of determining the RS codewords. This can be
done by computing a Fourier transform of length $N$ which can be
accomplished by $O(N \log_2(N))$ operations over the field
$\text{GF}(N)$. Addition over this field $\text{GF}(N)$ can be
implemented with $\log_2(N)$ binary operations and multiplication
can be implemented in $O((\log_2(N))^{\log_2(3)})$ binary operations.
Since one such codeword contains $N \log_2(N)$ bits, the claim
follows. A good reference for the spectral view of RS codes
is \cite{Bla84}.

\item {[The {\em decoding} complexity {\em per bit} is $O(\log_2(N))$
real operations and $O((\log_2(N))^{1+\log_2(3)})$ binary operations.]}
There are two components contributing to the decoding complexity.
First, we have to decode all  $N \log_2(N) \frac{2}{\epsilon}$ polar
blocks. Measured per bit this causes a complexity of $O(\log_2(N))$
real operations. The second operation is the erasure decoding of
the RS codes of length $N$ over $\text{GF}(N)$. This can be done
in $O(N (\log_2(N))^2)$ symbol operations (additions and multiplications) assuming that 
we preprocess and store $O(N \log_2(N))$ symbols, i.e., $O(N (\log_2(N))^2)$ bits, see \cite{Did09}.
Therefore the bit complexity of the processing is  $O(N (\log_2(N))^{2+\log_2(3)})$.
The claim follows since this takes care of 
of $O(N \log_2(N))$ bits.  \end{itemize}
\end{IEEEproof}

{\em Discussion:} At the decoder we need to know what indices belong
to the good set for the given BMS channel at hand. As was pointed
out in the introduction, this can be computed efficiently as shown
in \cite{TV, RHTT}.  Note that we only have to do this once every
time the channel changes.

For the boundaries by design we only use indices which are good for
all channels in BMS($C$). There is currently no efficient algorithm to compute
this set. But we can efficiently compute subsets, e.g., the set of
indices which is good for the channel which is the least degraded
with respect to the whole family $\text{BMS}(C)$, \cite{WHU13}.

It is easy to improve the error probability
substantially by using the RS code not only for erasure decoding
but also for error correction. We leave the details to the reader.

As a final remark. In the above complexity computation we list the
complexity of the polar decoding as the number of real operations.
In \cite{hamedthesis} it is shown how to accomplish the decoding in binary
operations if we allow a small gap in capacity.

\subsection{Variations on the Theme -- Two Channels}
Many variations on the basic construction are possible. Let us
briefly discuss one of them. Assume that we only have two
channels, i.e., ${\mathcal W}=\{\Ldens{a}, \Ldens{b}\}$ and that
rather than achieving the compound capacity $C({\mathcal W})$ we
just want to improve the achievable rate.

In this case we can construct a much shorter code. Let us quickly
explain.  Rather than stacking up $N$ polar blocks on top of each
other, we  only stack up $2^l$ for some $1 \leq l \leq n$.
Further, we only use a single extended staircase (rather than $n$).
Let us describe this scheme in some more detail.

\begin{figure}[ht!]
\begin{center}
\input{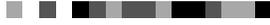}
\end{center}
\caption{The figure shows our graphical representation of one block
of length $16$ where the ``type'' of each index is indicated by a
particular shade of gray. White corresponds to bad for both channels, light gray
corresponds to bad for $\Ldens{a}$ but good for $\Ldens{b}$,
dark gray corresponds to the converse case, and black corresponds
to an index which is good for both channels. Note that we can
assume that number of light gray indices is equal to the number of
dark gray indices.}\label{fig:oneblocktwochannels} \end{figure}

Consider a basic polar block.  Each position $i$ in the basic block
of length $N$, $1 \leq i \leq N$, can have one of four possible types,
namely it can be in $\aset$ or not and it can be in $\bset$ or not.\footnote{In the sequel we write $\aset$ instead of $\aset_N$
in order to simplify our notation.}
Let us indicate this by shades of gray as shown in 
Figure~\ref{fig:oneblocktwochannels}.

Our construction is best seen visually. Take $2^l$ polar codes of
length $N$.  Visualize each such code as a row vector of height $N$
as in Figure~\ref{fig:oneblocktwochannels}. Place the $2^l$ row
vectors on top of each other but shift each copy one position further
to the right so that they visually form a ``staircase." To be
concrete, assume that the top-most copy is the one that is shifted
the furthest to the right. Further, take $k$ such basic units
(staircases) and place them next to each other in a contiguous
manner. The total blocklength of this construction is hence $k 2^l
N$.

\begin{figure}[ht!]
\begin{center}
\input{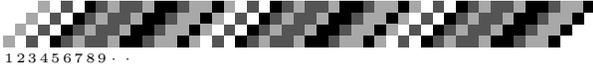}
\end{center}
\caption{
The construction for $N=16$, $l=2$, and $k=3$.
}\label{fig:constructiontwochannels} 
\end{figure}

Consider this construction for $N=16$, $l=2$, and $k=3$.
This is shown in Figure~\ref{fig:constructiontwochannels}.

To complete the construction we will now match pairs of indices
which appear in the same column. For the following explanation we
will refer to Figure~\ref{fig:constructiontwochannels}
for a concrete case.

Number the columns from left to right from $1$ to $kN+2^l-1$.  Note
that for each column in the range $2^l \leq i \leq k N$ we see $2^l$
distinct polar indices and that towards the boundary we still see
distinct polar indices but fewer.

For each column $i$ find a suitable matching.  For our example this
means the following. White boxes correspond to polar indices which
are bad for both channels. Freeze such boxes.  Black boxes correspond
to polar indices which are good for both channels.  To each such
box we associate an independent bit to be transmitted.  Finally,
try to match each light gray box (which corresponds to a polar index
which is good for the first but not good for the second channel)
with a dark gray box (which has the converse property).  Any index
that cannot be matched is set to frozen.

For position $1$ we see a single light gray box. Since we cannot
match it, we freeze it. For position $2$ we see a white box and a
light gray box. Again, freeze both.  For position $3$ we can associate
one bit to the black box and have to freeze the remaining two. The
first interesting position is position $7$ where we can in fact
match a light and dark gray box.

Note that any time we manage to match two boxes we win in rate
compared to standard polar codes.  Any time a (light or dark) gray
box cannot be matched we loose compared to the compound capacity.
Further note that if we had chosen $l=n$ then in each column in the
range $N \leq i \leq k N$ we would see {\em all} the polar indices
of one block. Hence, by our assumption that the number of light and
dark gray boxes in a block of length $N$ is equal, we can accomplish
a perfect matching in this case.  The only loss we still incur in
this case is at the boundaries where we only see part of the indices
of the whole block. But if we choose $k$ large then we can make
this loss as small as desired.

Finally note that this scheme can be decoded by a set of $2^l$
parallel successive decoders.  Each of the $2^l$
decoders runs on one row and the decoders are synchronized in the
sense that they decode at every instance the bits associated to a
particular column of the construction. Indices corresponding to
black boxes can be decoded under either channel condition. And for
matched boxes we can decode exactly one of the two in each case and
then determine the bit associated to the other box since we know
that it was a repetition. This way all decoders know the past history
under either channel condition and can hence proceed as in the
standard case.

{\em Discussion:} It is intuitive that already a small number of
copies suffices to reduce the compound gap significantly. Unfortunately
we do not have a bound on the convergence speed.  But it is tempting
to venture as guess that the gap decreases like one over the square
root of the number of copies.\footnote{Why is this a reasonable
guess? Assume that the various types were distributed over the
columns uniformly at random. In this case, if we look at a particular
column then the distribution of the types $(1, 0)$ and $(0, 1)$ are
both Bernoulli with equal mean. It follows that the expected number
of such types which cannot be matched decreases like one over the
square root of the number of copies. Our guess therefore stems from
assuming that for large blocklenghts the distribution effectively
looks like this random case.}


\section{Scheme II}\label{sec:finitesetone}
Let us now present an alternative construction which is capable of
achieving the compound capacity for a finite set of channels. We then show that this 
scheme is capable of achieving the compound capacity of any set of BMS channels. 
In the sequel we make no effort to optimize the various parameters.

\subsection{Polar Codes which are Good for Two
Channels}\label{sec:twochannels} 
Let us revisit the situation which we discussed in
Section~\ref{sec:basescheme}.  Recall that without essential loss
of generality we can assume that $C(\Ldens{a})=C(\Ldens{b})$ and
that $|\aset|=|\bset|$.

Consider the differences $\aset \setminus \bset=
\aset \cap \bset^c$ and $\bset \setminus \aset= \bset \cap \aset^c$,
where $(\cdot)^c$ denotes the complement of a set. Note that $|
\aset \setminus \bset | = | \bset \setminus \aset | $.

\begin{figure}[ht!] 
\begin{center} \includegraphics[width=3cm]{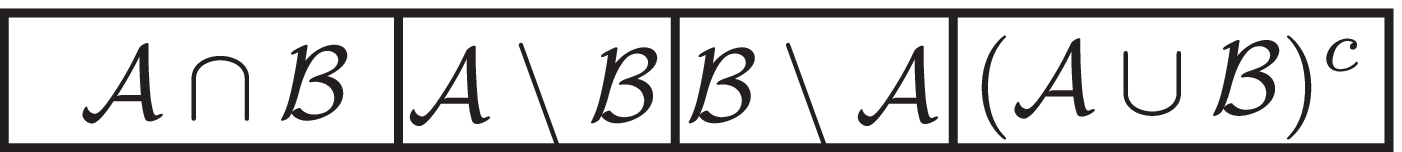}
\end{center} 
\caption{A graphical representation for the case of two channels.
The sets $\aset \cap \bset$, $\aset \setminus \bset$, $\bset \setminus
\aset$, and $(\aset \cup \bset)^c$ are indicated for one polar
block. For convenience of the representation the various sets are
drawn consecutively. The real such indices are of course spread out
over the whole block of length $N$.  }\label{fig:AB} \end{figure}

Let us represent the sets $\aset$ and $\bset$ as in Figure~\ref{fig:AB}.
The picture represents a polar block of length $N$ and shows
which indices belong to what ``type.''

\begin{definition}[Chaining Construction] \label{def:kchain}
Let $k \geq 2$. The {\em k-chain} of $\aset$ and $\bset$ is a code
of length $kN$ which consists of ``chaining'' together $k$ polar
blocks of length $N$ in the following manner.

In each of the $k$ blocks the set $\aset \cap \bset$ is an information
set.  Further, in block $i$, $1 \leq i < k$, the set $\aset \setminus
\bset$ is chained to the set $\bset \setminus \aset$ in block $(i+1)$
in the sense that the information is repeated in these two
sets (note that the two sets have the same cardinality). All other indices
are frozen.  Hence, the rate
of this construction is
\begin{equation} \label{ABkrate}
\frac{|\aset \cap \bset|+\frac{k-1}{k} |\aset \setminus \bset|}{N}.
\end{equation}
The scheme is visualized in Figure~\ref{fig:AB-half} for the case $k=3$.
\begin{figure}[ht!]
\begin{center} 
\includegraphics[width=8cm]{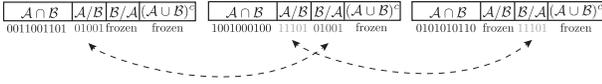}
\end{center}
\caption{The chaining construction with $k=3$. 
The dashed lines between two sets indicate that the information
in these two sets is repeated. }\label{fig:AB-half}
\end{figure}        
\EDe
\end{definition}

{\em Discussion:} Recall that if we were to use a standard polar
code for the compound scenario involving the channels $\Ldens{a}$ and $\Ldens{b}$
we could only transmit within the set $|\aset \cap \bset|$. This
results in a {\em rate-loss} of $\frac{|\aset \setminus \bset|}{N}$
compared to what we can achieve when transmitting over a single
channel. For the chaining construction the achievable rate on the
other hand can be made as close to $|\aset|/N$ as we want by choosing
$k$ sufficiently large.

\begin{example}
Let us go through the case with $k=3$ shown in Figure~\ref{fig:AB-half}
in more detail.

In block one (the left-most block in the figure) we put information
in the positions indexed by $\aset \cap \bset$ and $\aset \setminus
\bset$. The positions indexed by $\bset \setminus \aset$ as well
as $(\aset \cup \bset)^c$ are frozen and can be set to $0$.

In block two (the middle block in the figure) we put information
in the positions indexed by $\aset \cap \bset$ and $\aset \setminus
\bset$. In the positions indexed by $\bset \setminus \aset$ we
repeat the information which is in the positions indexed by  $\aset
\setminus \bset$ in block one.  The positions indexed by $(\aset
\cup \bset)^c$ are again frozen and can be set to $0$.

Finally, in block three (the right-most block in the figure) we put
information in the positions indexed by $\aset \cap \bset$.  In the
positions indexed by $\bset \setminus \aset$ we repeat the information
which is in the positions indexed by  $\aset \setminus \bset$ in
block two.  The positions indexed by $(\aset \cup \bset)^c$
are again frozen and can be set to $0$.
\EEx
\end{example}     

Let us now discuss how to decode this code.  The decoder sees the
received word and is aware of the channel which was used.  Since
the construction is symmetric we can assume without loss of generality
that it is the channel $\Ldens{a}$.  In this case the decoder can
decode block one (the left-most block in the figure) reliably.  This
is true since we only placed information in the sets $\aset \cap
\bset$ and $\aset \setminus \bset$, both of which are good for
channel $\Ldens{a}$.  All the other positions were frozen.  Once
block one has been decoded, we copy the information which was contained
in the set $\aset \setminus \bset$ to the position indexed by $\bset
\setminus \aset$ in block two. Now we can reliably decode block
two. Note that we have crucially used the fact that frozen positions
can contain any value as long as the value is known to the receiver.

We continue in this fashion. E.g., in the next step, copy the
information which was contained in block two in the positions indexed
by $\aset \setminus \bset$ to block three to the positions indexed
by $\bset \setminus \aset$.  Now we can reliably decode block three.
We go on with this scheme until we have reached block $k$.  If, on
the other hand, the information was transmitted on channel $\Ldens{b}$
we proceed in an equivalent fashion but start the decoding from the
right-most block.

What is the overall probability of error?  If we have $k$ blocks
then by a simple union bound the error probability is at most $k
\max\{P(\Ldens{a}), P(\Ldens{b})\}$.  Recall that if $k$ is large, then the common
rate in \eqref{ABkrate} tends to $|\aset|/N$, which we know can be
made as close to capacity as we desire by picking a sufficiently
large blocklength $N$ and a properly chosen index set $\aset$. Let
us summarize this discussion by formulating these observations as
lemmas.

\begin{lemma}[Chaining Construction is Good] \label{chaingood} Consider two BMS channels
of capacity $C$, $0 < C < 1$. Call the two channels $\Ldens{a}$ and $\Ldens{b}$. 
Assume that under the standard successive
decoding the good indices under channel $\Ldens{a}$ are $\aset$ and that the good indices under
channel $\Ldens{b}$ are $\bset$ and let $P(\Ldens{a})$ and $P(\Ldens{b})$ denote the respective single-channel
block error probabilities.

Then for each $k \geq 2$, the $k$-chain described in
Definition~\ref{def:kchain} has an error probability of at most $k
\max \{P(\Ldens{a}), P(\Ldens{b})\}$ for transmission over channel $\Ldens{a}$ as well for
transmission over channel $\Ldens{b}$ and a rate given in \eqref{ABkrate}.

If for a fixed $k$ we let $N$ tend to infinity then we will achieve
the rate $C(\{\Ldens{a} , \Ldens{b}\})-\frac{1}{k} \Delta(\Ldens{a} \cap \Ldens{b})$ and an arbitrarily
low error probability. If in addition we let $k$ tend to infinity
then we achieve the compound capacity $C(\{\Ldens{a}, \Ldens{b}\})$.
\end{lemma}

\subsection{A Polar View of Chains for Two Channels}\label{sec:polarview}
Let us now give a slightly different interpretation of the previous
construction. Rather than thinking of the previous construction as
a chain which is decoded either from the left to the right or vice
versa depending on which channel we use, let us observe that we can
construct from it a ``real'' polar block where we have a fixed
decoding order and use a standard successive decoder
according to this order.

Consider once again the situation depicted in Figure~\ref{fig:AB},
i.e., we have two channels called $\Ldens{a}$ and $\Ldens{b}$ and their respective
good sets are $\aset$ and $\bset$. Instead of constructing from
this a $2$-chain by matching up the indices of $\aset \setminus
\bset$ in one block with the indices of $\bset \setminus \aset$ in
the other block and by repeating the same information in these two
sets of indices, let us  combine the two blocks via a special
polar step where we match channels in a particular way. 

\begin{figure}[ht!] 
\begin{center} \includegraphics[width=9cm]{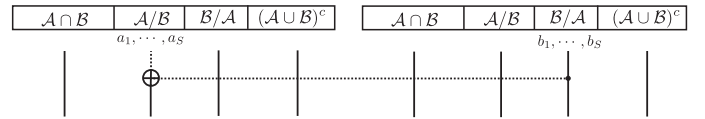}
\end{center} 
\caption{A graphical representations of the polar scheme for the case of two channels.}
\label{fig:polarscheme} \end{figure}

The scheme is shown in Figure~\ref{fig:polarscheme}. Whereas in the
$2$-chain the exact matching of the bits was immaterial, we will
now make a very specific choice.  Recall that by our convention the labels of
the polar indices are ordered according to the processing order of
the successive decoder. This means, if the indices $i$ go from $1$
to $N$, then we first process index $1$, then $2$, and so on. Let
$|\aset \setminus \bset| = |\bset \setminus \aset| = S$ and let
$\{a_1, \cdots, a_S\}$ denote the subset of $[1, N]$ which corresponds
to the set $\aset \setminus \bset$ and let $\{b_1, \cdots, b_S\}$
denote the equivalent for the set $\bset \setminus \aset$. Further,
we assume that these indices are ordered in a strictly increasing order. 

The polar transform which we use is indicated in
Figure~\ref{fig:polarscheme}.  Note that each of the two blocks
itself is a standard polar block of length $N$. The additional
polarization steps which are now performed are done solely for the
purpose of improving the compound capacity of the code.

The idea is to polarize the index $a_i$ of the first block with the
index $b_i$ of the second block. None of the other indices are
polarized but they are simply ``passed through'' unchanged. Associated
to this scheme we use the following processing order. We start with
index $1$ from block one. We go down the list of indices of block
one in the natural order until and including position $a_1-1$.  We
then process all the bits in block two just until and including
position $b_1-1$. We then process position $a_1$ of block one,
immediately followed by position $b_1$ from block two. At this point
we have performed our first ``matching'' by polarizing the first element
of $\aset \setminus \bset$ of block one with the first element of
$\bset \setminus \aset$ of block two. We now continue in exactly
the same fashion, processing elements of block one until and including
index $a_2-1$. The we process the positions of block two until and
including index $b_2-1$. We then process index $a_2$ of block one
followed immediately by processing index $b_2$ in block two and so
on.

Note that in this way we have created a new polar block 
of length $2 N$ and which has a specified processing order. Let us
introduce the following notation. We say that an index is of {\em
type} $(1, 0)$ if the index is contained in the set $\aset$ but not
in the set $\bset$.  This means, this is an index which is good for
the $\Ldens{a}$ channel but bad for the $\Ldens{b}$ channel. The equivalent 
definitions apply for the three remaining types, namely $(0, 0)$,
$(0, 1)$ and $(1, 1)$.

{\em The key observation} is that under successive decoding the
various indices have the following type:  An index which used to
have type $(0, 0)$ or $(1, 1)$ in either of the original blocks is
simply passed through by this construction and still has this type.
Let us clarify what we mean that it has this type. What we mean is
that if we {\em define} it to have this type then,  under successive
decoding, all indices which are declared to be of type $(1, 1)$
will have a small error probability. So in this sense these indices
are ``good.''

An index which used to have type $(0, 1)$ in the original block one
or type $(1, 0)$ in the original block two is also simply passed
through and maintains its type. 

But an index which used to be of type $(1, 0)$ in block one is now
polarized with an index of type $(0, 1)$ in block two and this
creates two new indices. The first one is of type $(0, 0)$ and the
second is of type $(1, 1)$. This means that we have converted two
indices, one which was good for $\Ldens{a}$ but bad for $\Ldens{b}$ and one which
was good for $\Ldens{b}$ but bad for $\Ldens{a}$ into two new indices where one is
bad for both and  one is good for both.

\begin{figure}[ht!]
\begin{center} 
\includegraphics[scale=.4]{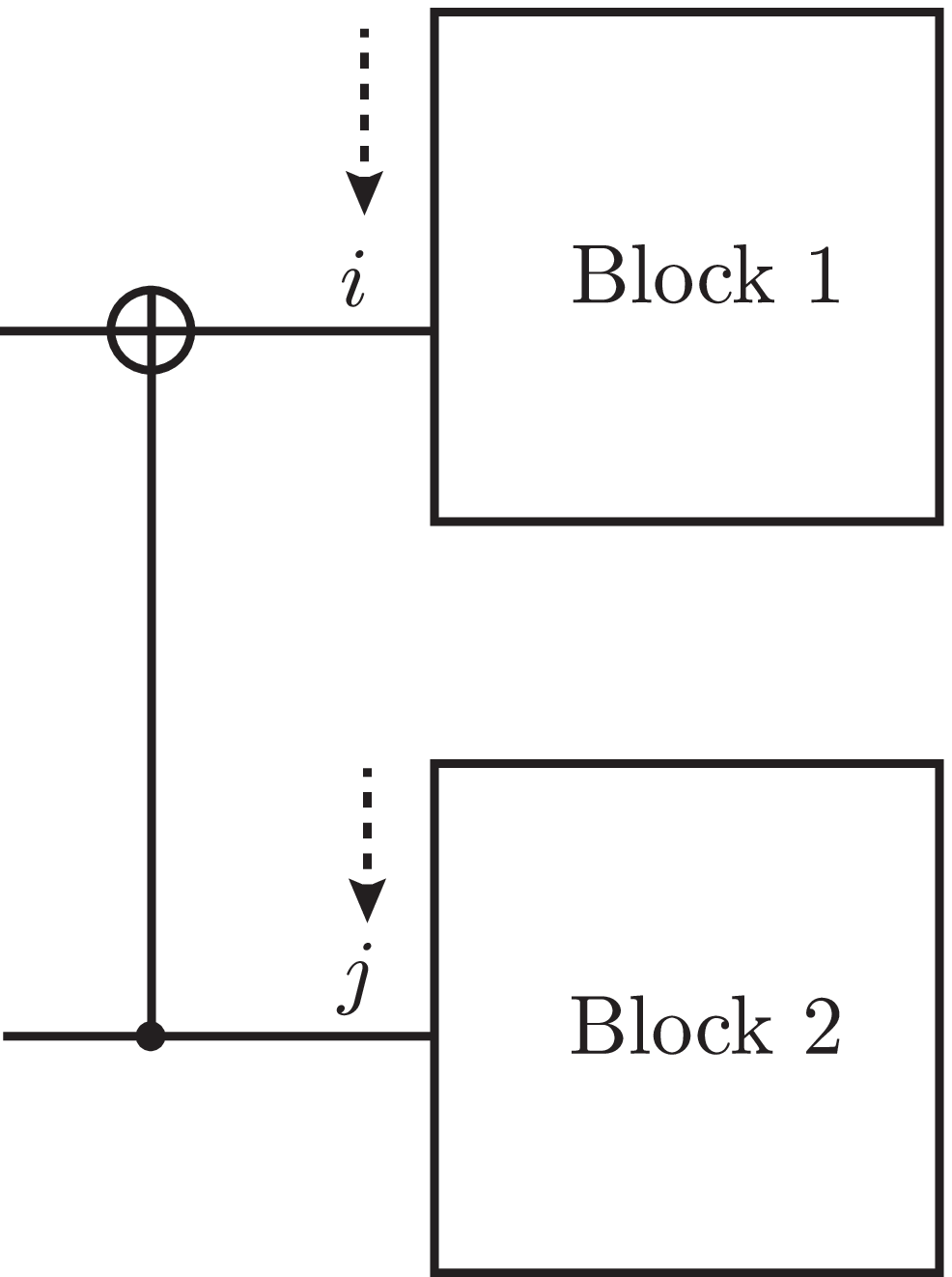}
\end{center}
\caption{How to combine two polar blocks to improve the compound capacity for two channels.}\label{fig:mainidea}
\end{figure}

The preceding paragraph encapsulates the main idea of the construction.
It is worth going over the two possible cases in more detail since
this is the main building block of our scheme. So consider a structure
like a standard polar transform but assume that you have two different
boxes as shown in Figure~\ref{fig:mainidea}.  The top box represents
a perfect channel in case the actual channel is $\Ldens{a}$ and a
completely useless channel in case it is $\Ldens{b}$. The box on
the bottom branch has the same property except that the roles of
$\Ldens{a}$ and $\Ldens{b}$ are exchanged. Now regardless which
channel is used, the top polar index (assuming uniform random
information along the bottom index) is a useless index, but once
we have processed and hence decided upon the top index the bottom
index is good for either of the two channels $\Ldens{a}$ and
$\Ldens{b}$ since either the information will flow along the top
branch or along the bottom branch.

If we use the so constructed polar block in a compound setting we
see that we have halved the gap $\Delta(\Ldens{a} \cap \Ldens{b})$.
Hence, the gap of the compound capacity of {\em this} polar block
compared to true compound capacity $C(\{\Ldens{a}, \Ldens{b}\})$
has been halved (assuming that we have sufficiently long blocks so
that for individual channels we essentially achieve capacity).

Even better. The block so constructed has exactly the same structural
properties as a standard polar block. It follows that we can recurse
this construction. If we perform $\kappa$ recursion steps then we
have created a block of length $2^\kappa N$ which has a block error
probability of at most $2^\kappa \max\{P(\Ldens{a}), P(\Ldens{b})\}$
and whose gap to the real compound capacity  $C(\{\Ldens{a},
\Ldens{b}\})$ has decreased by a factor $2^{-\kappa}$. In other
words, this construction is exactly as efficient as the chain
construction which we presented in the previous section.  But as
we will discuss in the next section, from this point of view the
generalization to more than two channels is immediate.

\subsection{Polar Codes Which are Good for a Finite Set of Channels}
Very little is needed to lift the previous construction for two
channels to a finite number of channels. Here is the general
recipe.  We start with two channels and a basic polar block of
length $N$, we then recurse $\kappa$ times until we have ``aligned''
essentially all good indices for these two channels. Those that are
still not aligned are thrown away.  We hence have a polar block of
length $2^\kappa N$ which is simultaneously good for two channels. We now
recurse on this block but with the aim of also aligning the good
indices for a third channel. We proceed in this fashion until we
have aligned the indices for all channels. 

Let us go over these steps a little bit more in detail.  Assume
that we have a finite set of channels, call them ${\mathcal W}=\{\Ldens{a}_1,
\cdots, \Ldens{a}_t\}$, all of capacity $C$, $0 < C <1$.

The construction is as follows.  We proceed recursively. We are given a
target gap of $\epsilon>0$ to the compound capacity $C({\mathcal
W})$ (here  we have $C({\mathcal
W})=C$).  Further, we are given a target block error probability
$P>0$.

We start with a standard polar block of length $N=2^n$, where $N$
will be chosen later on sufficiently large to fulfill the various
requirements.  Let $\Ldens{a}=\Ldens{a}_1$ and $\Ldens{b}=\Ldens{a}_2$.
Further, let $\aset=\aset_1$ and $\bset=\aset_2$.  Construct a polar
block which is good for channels $\Ldens{a}$ and $\Ldens{b}$ as
described in the previous section in the sense that the fraction
of indices which are good for channel $\Ldens{a}$ but not good for channel
$\Ldens{b}$ is at most $\epsilon/t$.  Let us assume that this takes
$\kappa_2$ recursions of the basic scheme.

Now let $\aset$ to be the set of indices which are jointly good for
channel $\Ldens{a}$ and $\Ldens{b}$ (this means in particular that
we ``throw away'' any indices which at this point are good for
only one channel; but we are assured that there are only few of
them so the rate-loss is minor). Further, set $\Ldens{b}=\Ldens{a}_3$
and $\bset=\aset_3$. Since our construction resulted in a
polar block we can recurse the construction, taking now as
our building block the block we previously constructed and which
has length $2^{\kappa_2} N$. In the second stage we recurse as many
times as are needed so that we incur an additional gap of at most
$\epsilon/t$. Assume that this takes $\kappa_3$ steps.

We proceed in this manner, adding always one channel at a time. In
this case we will have created a block which has the property that
the fraction of indices which are good for all the $t$ channels
simultaneously is at least $C-\epsilon$.  How many iterations do
we need? Consider one step of this process. A priori the gap to the
compound capacity is at most $\Delta(\cap_{i=1}^{t} \Ldens{a}_i)$.
For every extra polarization step we decrease this gap by a factor
of $\frac12$.  Therefore, we need at most $ \lceil \log_2
\frac{\Delta(\cap_{i=1}^{t} \Ldens{a}_i)t}{\epsilon} \rceil$ steps
so that the gap is no larger than $\epsilon/t$. If we have $t$ channels,
then we have $t-1$ such steps. We conclude that 
the sum of the required steps is at most
\begin{align*}
(t-1) \lceil \log_2 \frac{\Delta(\cap_{i=1}^{t} \Ldens{a}_i)t}{\epsilon} \rceil \leq
(t-1) (\log_2 \frac{\Delta(\cap_{i=1}^{t} \Ldens{a}_i)t}{\epsilon}+1),
\end{align*}
so that the total blocklength is at most $\bigl( \frac{2
\Delta(\cap_{i=1}^{t} \Ldens{a}_i) t}{\epsilon} \bigr)^{t-1} N$.

Note that since we have a fixed upper bound on the total number of
recursions (and this upper bound does not depend on the block length
$N$), it is possible to fix $N$ sufficiently large so that the final
block error probability is sufficiently small.

The above bounds are quite pessimistic and an actual construction
is likely to have significantly better parameters. In particular,
once we have constructed a polar code which is good for several
channels, it is likely that for any further channel we add we only
need a few recursions. One would therefore assume that the pre-factor
which is currently  $\bigl( \frac{2 \Delta(\cap_{i=1}^{t} \Ldens{a}_i)
t}{\epsilon} \bigr)^{t-1}$ is in reality much smaller.

\subsection{Compactness of the Space of BMS Channels}\label{sec:compactness}
It remains to transition from a finite set of BMS channels to the
whole set of BMS channels, lets say of capacity $C$, call this set
$\text{BMS}(C)$. The crucial observation in this respect is that
(i) $\text{BMS}(C)$ is compact, and that (ii) we can modify the
finite set of representatives implied by the compactness so that
all channels in $\text{BMS}(C)$ are upgraded with respect to these
representatives.

\begin{lemma}[Construction of a Dominating Set]\label{lem:finitedominatingset}
Let $\text{BMS}(C)$ denote the set of BMS channels of capacity at
least $C$.  Let $\epsilon>0$. Then we can explicitly construct a set 
of channels of cardinality $K(\epsilon)$, denote
this set by $\{\Ldens{c}_i\}_{i=1}^{K(\epsilon)}$, with the following
two properties:
\begin{itemize}
\item[(i)] $C(\Ldens{c}_i) \geq C-\epsilon$ for all $1 \leq i \leq K(\epsilon)$. 
\item [(ii)] For any
$\Ldens{c} \in \text{BMS}(C)$, there exists at least one $i$, $1
\leq i \leq K(\epsilon)$, so that $\Ldens{c}_i \prec \Ldens{c}$.
\end{itemize}
In words, every channel in $\text{BMS}(C)$ is upgraded with respect
to at least one channel in $\{\Ldens{c}_i\}_{i=1}^{K(\epsilon)}$,
and every $\Ldens{c}_i$ has capacity at least equal to $C-\epsilon$.
We have the bound $K(\epsilon) \leq \binom{2A(\epsilon)}{A(\epsilon)} =
\frac{4^{A(\epsilon)}}{\sqrt{A(\epsilon) \pi}}(1+O(1/A(\epsilon)))$ where 
\begin{align*}
A(\epsilon) =  \lceil \frac{9}{8 h_2^{-1}(\epsilon)^2} \rceil.
\end{align*}
\end{lemma} 
\begin{IEEEproof}
The proof uses the machinery developed in \cite{RiU08,KRU12}.  In
brief, every BMS channel can be represented by a probability density
on the unit interval $[0,1]$ (this is sometimes called the
$|D|$-representation, see \cite{RiU08}). Consider a BMS channel
$\Ldens{c}$. By some abuse of notation we let $\Ldens{c}(x)$ denote
the density on $[0,1]$ which represents $\Ldens{c}$. The capacity
of $\Ldens{c}$ can be computed from $\Ldens{c}(x)$ via  the integral
\begin{equation*}
C(\Ldens{c})=1-\int_{0}^1 h_2\Bigl( \frac{1-x}{2} \Bigr) \Ldens{c}(x) \text{d}x,
\end{equation*}
where $h_2(x)$ denotes the binary entropy function, $h_2(x) = -x
\log_2(x) -(1-x) \log_2(1-x)$.  Further, as in \cite{RiU08}, we
equip the space $[0,1]$ with the Wasserstein metric which we denote
by $d(\cdot,\cdot)$.

Let us first show how to find a set of BMS channels, denote them
by $\{\Ldens{e}_i\}_{i=1}^{K(\epsilon)}$, such that for any $\Ldens{c}
\in \text{BMS}(C)$, there exists at least one $i$, $1 \leq i \leq
K(\epsilon)$, so that
\begin{align*}
d(\Ldens{c} , \Ldens{e}_i)< \frac{4}{9} h_2^{-1}(\epsilon)^2.
\end{align*}
Define the discrete alphabet ${\mathcal X} = \{x_0, x_1, \cdots,
x_{T-1}, x_T\}$, where $x_i = \frac{i}{T}$, $0 \leq i \leq T$, and
where $T=\lceil \frac{9}{2  h_2^{-1}(\epsilon)^2} \rceil$. In other words,
${\mathcal X}$ consists of $T+1$ points equally spaced in the unit
interval $[0, 1]$.

Consider the space of all densities that have the following form
\begin{equation}\label{equ:capformula}
\sum_{i=0}^T p_i \delta(x-x_i),
\end{equation} 
where $\sum_i p_i =1$ and $p_i \in {\mathcal X}$. Let us denote
this space by $\text{BMS}_T$.  A computation shows that this set has cardinality
$\binom{2 T}{T}$. From the properties of the Wasserstein distance
we know that for any BMS channel $\Ldens{c}$, there exists an
$\Ldens{e} \in \text{BMS}_T$ such that 
\begin{equation} \label{d<2/T}
d(\Ldens{c},\Ldens{e}) \leq \frac{2}{T} \leq  \frac{4}{9} h_2^{-1}(\epsilon)^2.  
\end{equation}
We construct now from each $\Ldens{e}_i$, $1\leq i
\leq K(\epsilon)$, another density $\Ldens{c}_i$ such that
\begin{itemize}
\item[(i)] $C(\Ldens{c}_i) \geq C(\Ldens{c}_i) - \epsilon$.
\item[(ii)] For any $\Ldens{c} \in \text{BMS}(C)$ there exists and $i$ so that $\Ldens{c}_i \prec \Ldens{c}$.
\end{itemize}
It is shown in  Corollary 43 in \cite{KRU12} how to modify the set
$\{\Ldens{e}_i\}_{i=1}^{K(\epsilon)}$ into the set
$\{\Ldens{c}_i\}_{i=1}^{K(\epsilon)}$ to ensure the degradedness
condition and so that the Wasserstein distance, assuming it was
$\delta$ beforehand, is at most $3 \sqrt{\delta}$ afterwards. But a
Wasserstein distance of $3 \sqrt{\delta}$ implies a loss in capacity
of at most  $h_2\Bigl( \frac32 \sqrt{\delta} \Bigr)$.
\end{IEEEproof}

\section{Discussion and Open Questions}
Let us quickly discuss some interesting open questions.  Although
we have proposed two solutions which solve the compound capacity
problem at low computational cost, both solutions require a significant
increase in blocklength compared to standard polar codes.  It is
therefore interesting to see if we can find variations of the
proposed solutions, or perhaps different solutions that do not
suffer from the same problem.

Let us first consider scheme I. Recall that we payed in blocklength
a factor $N \log_2(N)$ compared to standard polar codes. The factor
$N$ was due to the fact that we stacked up $N$ polar blocks on top
of each other.  This simplified the analysis considerably since in
this case we know that the number of good indices within one column
is essentially independent of the channel and very close to $NC$.
It is tempting to conjecture that a much smaller number would
suffice. The crucial property which one needs is that the number
of good indices within this set does not vary significantly as a
function of the channel. We leave it as an interesting open question
if the column  height can be chosen significantly smaller than $N$.

Let us now look at scheme II. In this case one significant source
of inefficiency of the scheme stems from the fact that we first
performed polarization steps to achieve a good polarization for
single channels and then in a second step performed polarization
steps to achieve universality. It is tempting to conjecture in this
case that the two operations can be performed together and that in
fact for the second stage we do not need to explicitly polarize for
each single channel but that we can achieve universality in a more
direct and natural way by modifying the basic construction. I.e.,
one could imagine that for the basic construction we decide at each
step not to polarize all indices but only a fraction and that the
mix the indices which are polarized. The price we might have to pay
is a somewhat slower polarization but we might be able to achieve
universality directly. The investigation of such ideas is slated
for future research.

Finally, let us point out that such universal polar schemes might also 
be useful in other information theoretic scenarios (see for example 
\cite{HOF10, MELK13, BB13, ASOS13}).

\section{acknowledgment}
We would like to thank Eren \c{S}a\c{s}o\u{g}lu and Lele Wang for lively,
stimulating, and fruitful discussions.  This work was supported by
EC grant FP7-265496, ``STAMINA.''

\bibliographystyle{IEEEtran}
\bibliography{lth,lthpub}


\end{document}